%% file: main.tex
\documentclass[12pt]{article}
\usepackage{times}
\usepackage{geometry}
\usepackage{graphicx}

\usepackage[square,numbers]{natbib}

\usepackage{multirow}
\usepackage{tabularx}
\usepackage{booktabs}
\usepackage{float}
\usepackage[outercaption]{sidecap} 
\usepackage{threeparttable}

\usepackage{hyperref}
\hypersetup{
  colorlinks = true,
  allcolors = blue,
  linktocpage = true
}

\begin{document}


\begin{center}
{\bf\LARGE Astro2020 APC White Paper}

\vspace{2mm}
{\bf\Large Project: The Simons Observatory}

\end{center}

{\bf Project area:} Radio, Millimeter, and Submillimeter Observations from the Ground
 
 {\bf Corresponding author email:} Adrian.Lee@berkeley.edu


\vspace{10mm}
\thispagestyle{empty}

\begin{center}
{\small
\input{authors}

}
\vspace{8mm}
\end{center}
\input{address}


\clearpage
\setcounter{page}{1}


\vspace{-20mm}
\section{Executive Summary}

The Simons Observatory (SO) is a ground-based cosmic microwave background (CMB) experiment sited on Cerro Toco in the Atacama Desert in Chile that promises to provide breakthrough discoveries in fundamental physics, cosmology, and astrophysics. 
Supported by the Simons Foundation, the Heising-Simons Foundation, and with contributions from collaborating institutions,
SO will see first light in 2021 and start a five year survey in 2022. SO has 287 collaborators from 12 countries and 53 institutions, including 85 students and 90 postdocs.

The SO experiment in its currently funded form (`SO-Nominal') consists of three 0.4~m Small Aperture Telescopes (SATs)  and one 6~m Large Aperture Telescope (LAT).   Optimized for minimizing systematic errors in polarization measurements at large angular scales,
the SATs will perform a deep, degree-scale survey of 10\% of the sky  to search for the signature of primordial gravitational waves.
The LAT will survey 40\% of the sky with arc-minute resolution.  These observations will measure (or limit) the sum of neutrino masses, search for light relics, 
measure the early behavior of Dark Energy, and refine our understanding of the intergalactic medium, clusters and the role of feedback in galaxy formation.

With up to ten times the sensitivity and five times the angular resolution of the {\it Planck} satellite, and roughly an order of magnitude increase in mapping speed over currently operating (``Stage 3") experiments, SO will measure the CMB temperature and polarization fluctuations to exquisite precision in six frequency bands from 27 to 280 GHz. SO will rapidly advance CMB science while informing the design of future observatories such as CMB-S4. Construction of SO-Nominal is fully funded, and operations and data analysis are funded for part of the planned five-year observations. We will seek federal funding to complete the observations and analysis of SO-Nominal, at the \$25M level. 
The SO has a low risk and cost efficient upgrade path -- the 6~m LAT can accommodate almost twice the baseline number of detectors and the SATs can be duplicated at low cost.  
We will seek funding at the \$75M level for an expansion of the SO (`SO-Enhanced') that fills the remaining focal plane in the LAT, adds three SATs, and extends operations by five years, substantially improving our science return. By this time SO may be operating as part of the larger CMB-S4 project.

This white paper summarizes and extends material presented in \cite{so_forecast:2019}, which describes the science goals of SO-Nominal, and \cite{galitzki/etal:2018,gallardo/etal:2018,orlowski-scherer/etal:2018,dicker/etal:2018,zhu/etal:2018,coppi/etal:2018,hill/etal:2018,parshley/etal:2018} which describe the instrument design.  

\par \noindent

\vspace{-.2in}
\section{Key Science Goals and Objectives}
\label{sec:science}

\input science.tex

\input experiment.tex

\label{sec:enhance-subsec}

\section{Simons Observatory and Beyond\label{sec:enhance}}

\vspace{-.1in}
\subsection{Enhanced Simons Observatory: science goals and technical overview}

While SO-Nominal will significantly advance our understanding of cosmology beyond currently operating instruments, there are extensions that will enhance its capabilities.  Table 2 shows highlights of the potential science yield of SO-Enhanced that has the following upgrades: (a) filling the remaining six optics tubes in the LAT;
(b) adding three additional SATs; and (c) five additional years of operations. 

 This LAT enhancement will improve the signal-to-noise in SO's lensing maps by $\sim 70\%$, with most of the signal
 coming from polarization measurements, expected to be free of many foreground-induced errors that may limit temperature maps (Table 2). The SZ measurements will improve significantly, increasing the number of clusters from $\sim$20,000 to $\sim$30,000. In combination with LiteBIRD's $\tau$ measurement, SO-Enhanced targets a 4$\sigma$ detection of neutrino mass even for the minimal mass, using either CMB lensing or SZ cluster counts combined with large-scale-structure measurements.

With the SAT enhancement, SO could achieve a $1 \sigma$ limit on $r$ of 0.001 (Table 2). This is based on our current best estimates of foregrounds. There is a strong theoretical motivation to reach this level, with certain large-field plateau inflationary models predicting $r \approx 0.003$ \citep[e.g.,][]{starobinsky:1980}.

\begin{table*}[htb!]
\centering
\caption[Simons Observatory Surveys]{Highlights of the SO-Enhanced science 
} 
\begin{tabular}{ l | cc}
\hline
\hline
&  SO-Nominal  & SO-Enhanced  \\
  &   (Goal) &  (Goal) \\
\hline
\hline
{\bf Lensing and SZ (LAT)} &&\\
\hline
\quad Minimal neutrino mass detection ($\mnu$=0.06~eV)\textsuperscript{a}   &  3$\sigma$ &  4$\sigma$  \\

\quad Lensing  detection (polarization-only)& 160$\sigma$ (110$\sigma$)&220$\sigma$ (180$\sigma$)\\
\quad Number of SZ clusters  & 20000 & 33000  \\
\quad Kinematic SZ detection (DESI cross-correlation) & 190$\sigma$ & 240$\sigma$ \\
\quad Measurement of Optical Depth from kSZ, $\sigma(\tau)$
\textsuperscript{b} & 0.007 & 0.003 \\
\hline
{\bf Primordial polarization (SATs)}  &&\\
\hline
\quad Tensor-to-scalar  ratio &$\sigma(r)=0.002$ & $\sigma(r)=0.001$\textsuperscript{c} \\
\hline
\hline
\end{tabular}
\begin{tablenotes}
\item \textsuperscript{a} This can be derived from {\it either} CMB lensing {\it or} SZ clusters, each in combination with LSS. Here we assume a cosmic-variance optical depth measurement from LiteBIRD.
\item \textsuperscript{b} This uses the kSZ-induced 4-point function, a new method to study patchy reionization \cite{ferraro/smith:2018}. This forecast does not include astrophysical systematic uncertainties.
\item \textsuperscript{c} The SO-Nominal case projects $\sigma(r)=0.0018$ (2 s.f.) for a residual lensing power of $A_L=0.5$. For SO-Enhanced with $A_L=0.5$ we project $\sigma(r)=0.0012$; for $A_L=0.3$ we project $\sigma(r)=0.0009$. 
\end{tablenotes}
\label{tab:enhance}
\vskip -0.3in
\end{table*}


\subsection{Relationship to CMB-S4}

The SO is an important step toward the science goals of CMB-S4.  The timing is such that as CMB-S4 moves forward, SO technology development, infrastructure and analysis pipelines could be partially or fully incorporated into it.  In the meantime, investment in SO-Nominal and SO-Enhanced will dramatically accelerate accumulated advances in the field.  In particular, SO's highly sensitive multi-wavelength maps should help inform CMB-S4's frequency coverage and mitigation of systematic uncertainties.  SO members comprise 60\% of the CMB-S4 collaboration.  There is interest in the SO collaboration to move SO forward as fast as possible while supporting CMB-S4 as much as possible.  
A combination of public and private investment will make this synergistic plan possible. 

\vspace{-.2in}
\section{Schedule and cost estimates}
\label{sec:funding}

The SO project is well into the construction phase.  The first SAT and LATR cryostats have been delivered and are undergoing cryogenic verification.  The LAT and SATs are under construction and will be complete by 2021 and 2020 respectively.  Production of the detectors and readout will begin at the end of 2019. The SO schedule has first light for the first of the three SATs at the beginning of 2021 and full operations of the observatory in 2022.  The SO-Nominal goals outlined in Table \ref{tab:enhance} will be achieved during the first five years of observations from 2022 through 2027.  SO-Enhanced instrument upgrades will be phased into the SO infrastructure and will be completed by 2027.  The SO-Enhanced goals will be achieved after an additional five years of observing from 2027 through 2032: by this time SO may be operating as part of the larger CMB-S4 project.

The SO project budget is given in Table \ref{tab:budget}.  A total of \$124M in commitments have been made towards achieving the SO-Nominal goals with an additional \$25M required for observations and analysis.  All of the SO-Enhanced instrumentation will be copies of existing instrumentation with no required technology development, hence the costs are well understood. The observation and analysis cost estimates are based on 20 years of operational experience at the site in Chile as well as scaled (from ACT) estimates of the analysis effort. All costs are in FY19 dollars. 

It is possible that international support could be incorporated into the SO-Enhanced plan.  
For example, there are pending proposals to the UK Science and Technology Facilities Council, and to the Japanese large-scale project prioritization process, that could fund additional SATs.


\begin{table*}[htb!]
\centering
\caption[Simons Observatory Budget]{The SO-Nominal and SO-Enhanced Budgets. } \small
\begin{tabular}{ l | c c c}
\hline
\hline
  &  2017-2022 & 2022-2027 & 2027-2032\\
\hline
\hline
{\bf SO-Nominal} &&\\
\hline
\quad Simons and Heising-Simons Foundation Project Funding & \$60M \\

\quad Institutional Commitments (cash) & \$10M \\
\quad Institutional Commitments (reduced overhead - est.) & \$10M \\
\quad Institutional Commitments (in kind - est.) \\
\quad \quad 30 graduate students per year & \$9M &  & \\
\quad \quad 36 postdocs per year & \$15M &  &  \\
\quad Simons Foundation - Observations/Logistics && \$10M\\
\quad Simons Foundation - Pipeline/Analysis && \$10M \\
\quad Federal Support - Observations/Logistics & & {\bf \$15M} \\
\quad Federal Support - Pipeline/Analysis/Community && {\bf \$10M}  \\
\hline
{\bf SO-Enhanced}  &&\\
\hline
\quad Federal Support - Three SATs & & {\bf \$15M} \\
\quad Federal Support - Six LATR Optics tubes && {\bf \$12M} \\
\quad Federal Support - Observations/Logistics &&& {\bf \$30M}\\
\quad Federal Support - Pipeline/Analysis/Community &&& {\bf\$20M}\\
\hline
\hline
\end{tabular}
\label{tab:budget}
\end{table*}

\clearpage
\small{

\bibliography{so}  
}

\end{document}

%% file: authors.tex
The~Simons~Observatory~Collaboration:
Maximilian~H.~Abitbol$^{1}$,
Shunsuke~Adachi$^{2}$,
Peter~Ade$^{3}$,
James~Aguirre$^{4}$,
Zeeshan~Ahmed$^{5,6}$,
Simone~Aiola$^{7}$,
Aamir~Ali$^{8}$,
David~Alonso$^{1}$,
Marcelo~A.~Alvarez$^{8,9}$,
Kam~Arnold$^{10}$
% ...

%% file: address.tex

\noindent {\scriptsize{%
{$^{1}$~University of Oxford, Denys Wilkinson Building, Keble Road, Oxford OX1 3RH, UK}\\ 
{$^{2}$~Department of Physics, Faculty of Science, Kyoto University, Kyoto 606, Japan}\\ 
{$^{3}$~School of Physics and Astronomy, Cardiff University, The Parade, Cardiff, CF24 3AA, UK}\\ 
{$^{4}$~Department of Physics and Astronomy, University of Pennsylvania, 209 South 33rd Street, Philadelphia, PA, USA 19104}\\ 
{$^{5}$~Kavli Institute for Particle Astrophysics and Cosmology, Menlo Park, CA 94025}\\ 
{$^{6}$~SLAC National Accelerator Laboratory, Menlo Park, CA 94025}\\ 
{$^{7}$~Center for Computational Astrophysics, Flatiron Institute, 162 5th Avenue, New York, NY 10010, USA}\\ 
{$^{8}$~Department of Physics, University of California, Berkeley, CA, USA 94720}\\ 
{$^{9}$~Computational Cosmology Center, Lawrence Berkeley National Laboratory, Berkeley, CA 94720, USA}\\ 
{$^{10}$~Department of Physics, University of California San Diego, CA, 92093 USA}\\ 
{$^{11}$~Physics Division, Lawrence Berkeley National Laboratory, Berkeley, CA 94720, USA}\\ 
{$^{12}$~Kavli Institute for The Physics and Mathematics of The Universe (WPI), The University of Tokyo, Kashiwa, 277- 8583, Japan}\\ 
{$^{13}$~Joseph Henry Laboratories of Physics, Jadwin Hall, Princeton University, Princeton, NJ, USA 08544}\\ 
{$^{14}$~NIST Quantum Sensors Group, 325 Broadway Mailcode  687.08, Boulder, CO, USA 80305}\\ 
{$^{15}$~Department of Physics and Astronomy, Rutgers,  The State University of New Jersey, Piscataway, NJ USA 08854-8019}\\ 
{$^{16}$~International School for Advanced Studies (SISSA), Via Bonomea 265, 34136, Trieste, Italy}\\ 
{$^{17}$~INFN, Sezione di Trieste, Padriciano, 99, 34149 Trieste, Italy}\\ 
{$^{18}$~Department of Physics, University of Michigan, Ann Arbor, USA 48103}\\ 
{$^{19}$~Kavli Institute for Cosmology Cambridge, Madingley Road, Cambridge CB3 0HA, UK}\\ 
{$^{20}$~Institute of Astronomy, Madingley Road, Cambridge CB3 0HA, UK}\\ 
{$^{21}$~Department of Physics and Astronomy, University of New Mexico, Albuquerque, NM 87131, USA}\\ 
{$^{22}$~Department of Astronomy, Cornell University, Ithaca, NY, USA 14853}\\ 
{$^{23}$~Jodrell Bank Centre for Astrophysics, School of Physics and Astronomy, University of Manchester, Manchester, UK}\\ 
{$^{24}$~AstroParticule et Cosmologie, Univ Paris Diderot, CNRS/IN2P3, CEA/Irfu, Obs de Paris, Sorbonne Paris Cit\'e, France}\\ 
{$^{25}$~DAMTP, Centre for Mathematical Sciences, University of Cambridge, Wilberforce Road, Cambridge CB3 OWA, UK}\\ 
{$^{26}$~Department of Physics, Yale University, New Haven, CT 06520, USA}\\ 
{$^{27}$~School of Physics, University of Melbourne, Parkville, VIC 3010, Australia}\\ 
{$^{28}$~Instituto de Astrof\'isica and Centro de Astro-Ingenier\'ia, Facultad de F\`isica, Pontificia Universidad Cat\'olica de Chile, Av. Vicu\~na Mackenna 4860, 7820436 Macul, Santiago, Chile}\\ 
{$^{29}$~Canadian Institute for Theoretical Astrophysics, University of Toronto, 60 St.~George St., Toronto, ON M5S 3H8, Canada}\\ 
{$^{30}$~School of Electrical, Computer, and Energy Engineering, Arizona State University, Tempe, AZ USA}\\ 
{$^{31}$~Department of Physics \& Astronomy, University of Sussex, Brighton BN1 9QH, UK}\\ 
{$^{32}$~Department of Physics, York University, 157, 161 Campus Walk, North York, ON}\\ 
{$^{33}$~Department of Physics, Cornell University, Ithaca, NY, USA 14853}\\ 
{$^{34}$~Institute for Advanced Study, 1 Einstein Dr, Princeton, NJ 08540}\\ 
{$^{35}$~Imperial College London, Blackett Laboratory, SW7 2AZ UK}\\ 
{$^{36}$~Department of Applied and Engineering Physics, Cornell University, Ithaca, NY, USA 14853}\\ 
{$^{37}$~Astrophysics and Cosmology Research Unit, School of Mathematics, Statistics and Computer Science, University of KwaZulu-Natal, Durban 4041, South Africa}\\ 
{$^{38}$~Department of Physics, Stanford University, Stanford, CA 94305}\\ 
{$^{39}$~The Oskar Klein Centre for Cosmoparticle Physics, Department of Physics, Stockholm University, AlbaNova, SE-106 91 Stockholm, Sweden}\\ 
{$^{40}$~Department of Astrophysical Sciences, Peyton Hall,  Princeton University, Princeton, NJ,  USA 08544}\\ 
{$^{41}$~Simon Fraser University, 8888 University Dr, Burnaby, BC V5A 1S6, Canada}\\ 
{$^{42}$~University of Southern California. Department of Physics and Astronomy, 825 Bloom Walk ACB 439, Los Angeles, CA 90089-0484}\\ 
{$^{43}$~School of Earth and Space Exploration, Arizona State University, Tempe, AZ, USA 85287}\\ 
{$^{44}$~University of Arizona, 933 N Cherry Ave, Tucson, AZ 85719}\\ 
{$^{45}$~Department of Physics and Astronomy, University of Pittsburgh, Pittsburgh, PA, USA 15260}\\ 
{$^{46}$~Department of Physics and Astronomy, University of British Columbia, Vancouver, BC, Canada V6T 1Z1}\\ 
{$^{47}$~Physics and Astronomy Department, Stony Brook University, Stony Brook, NY USA 11794}\\ 
{$^{48}$~High Energy Accelerator Research Organization (KEK), Tsukuba, Ibaraki 305-0801, Japan}\\ 
{$^{49}$~Department of Astronomy and Astrophysics, The Pennsylvania State University, University Park, PA 16802}\\ 
{$^{50}$~Institute for Gravitation and the Cosmos, The Pennsylvania State University, University Park, PA 16802}\\ 
{$^{51}$~Astronomical Institute, Graduate School of Science, Tohoku University, Sendai, 980-8578, Japan}\\ 
{$^{52}$~Department of Physics, Florida State University, Tallahassee FL, USA 32306}\\ 
{$^{53}$~Department of Astronomy and Astrophysics, University of Toronto, 50 St.~George St., Toronto, ON M5S 3H4, Canada}\\ 
{$^{54}$~Dunlap Institute for Astronomy and Astrophysics, University of Toronto, 50 St.~George St., Toronto, ON M5S 3H4, Canada}\\ 
{$^{55}$~School of Physics and Astronomy, Sun Yat-Sen University, 135 Xingang Xi Road, Guangzhou, China}\\ 
{$^{56}$~Perimeter Institute 31 Caroline Street North, Waterloo, Ontario, Canada, N2L 2Y5}\\ 
{$^{57}$~Department of Physics, University of Tokyo, Tokyo 113-0033, Japan}\\ 
{$^{58}$~Kavli Institute for the Physics and Mathematics of the Universe (WPI), Berkeley Satellite, the University of California, Berkeley 94720, USA}\\ 
{$^{59}$~Research Center for the Early Universe, School of Science, The University of Tokyo, Bunkyo-ku, Tokyo 113-0033, Japan}\\ 
{$^{60}$~Laboratoire de l'Acc\'el\'erateur Lin\'eaire, Univ. Paris-Sud, CNRS/IN2P3, Universit\'e Paris-Saclay, Orsay, France}\\ 
{$^{61}$~Van Swinderen Institute for Particle Physics and Gravity, University of Groningen, Nijenborgh 4, 9747 AG Groningen, The Netherlands}\\ 
{$^{62}$~Department of Physics, Southern Methodist University, 3215 Daniel Ave. Dallas, Texas 75275-0175}\\ 
{$^{63}$~Department of Physics, University of Milano - Bicocca, Piazza della Scienza, 3, 20126 Milano, Italy}\\ 
{$^{64}$~University of Ferrara, Via Savonarola, 9, 44121 Ferrara FE, Italy}\\ 
{$^{65}$~Department of Physics and Astronomy, Haverford College,Haverford, PA, USA 19041}\\ 
{$^{66}$~Raymond and Beverly Sackler School of Physics and Astronomy, Tel Aviv University, P.O. Box 39040, Tel Aviv 6997801, Israel}\\ 
{$^{67}$~Department of Physics and Engineering, West Chester University of Pennsylvania, 720 S. Church St., West Chester, PA 19383}\\ 
{$^{68}$~McWilliams Center for Cosmology, Department of Physics, Carnegie Mellon University, 5000 Forbes Avenue, Pittsburgh, PA 15213, USA}\\ 
{$^{69}$~Department of Physics and Astronomy, University of California Los Angeles, 475 Portola Plaza, Los Angeles, CA 9009}\\ 
{$^{70}$~NASA/Goddard Space Flight Center, Greenbelt, MD, USA 20771}\\ 
{$^{71}$~San Diego Supercomputer Center, University of California San Diego, CA, 92093 USA}\ 
 }}

%% file: science.tex
\newcommand{\map}{{\sl WMAP}}
\newcommand{\planck}{{\it Planck}}
\newcommand{\wmap}{{\it WMAP}}
\newcommand{\cobe}{{\sl COBE}}
\newcommand{\ba}{\begin{eqnarray}}
\newcommand{\ea}{\end{eqnarray}}
\newcommand{\n}      {{\bf{n}}}
\newcommand{\EV}[1]  {\langle#1\rangle}
\newcommand{\lmax}   {l_{\rm max}}
\newcommand{\mmax}   {m_{\rm max}}
\newcommand{\uKsq}   {\mbox{$\mu{\rm K}^2$}}
\newcommand{\uKrs}   {\mbox{$\mu{\rm K}\sqrt{\rm s}$}}
\newcommand{\dg}     {\mbox{$^{\circ}$}}
\newcommand{\lsim}   {\mbox{$_<\atop^{\sim}$}}
\newcommand{\gsim}   {\mbox{$_>\atop^{\sim}$}}
\newcommand{\lt}     {\mbox{$<$}}
\newcommand{\gt}     {\mbox{$>$}}
\newcommand{\order}  {{\cal O}}
\newcommand{\JJ}     {{\cal J}}
\newcommand{\<}      {\langle}
\renewcommand{\>}    {\rangle}
\newcommand{\amin}   {\mbox{$^\prime\ $}}
\newcommand{\asec}   {\mbox{$^{\prime\prime}\ $}}
\newcommand{\ddeg}   {\mbox{${\rlap.}^\circ$}}

\newcommand{\ylm}[2] {{Y_{#1#2}}}
\newcommand{\ylmcc}[2]{{Y^{*}_{#1#2}}}
\newcommand{\fnlKS}  {f_{NL}^{\rm local}}
\newcommand{\fnleq}  {f_{NL}^{\rm equil}}

\newcommand  \cm     {{\rm \,cm}}
\newcommand  \gtsim  {\lower.5ex\hbox{$\; \buildrel > \over \sim \;$}}
\newcommand  \ltsim  {\lower.5ex\hbox{$\; \buildrel < \over \sim \;$}}
\newcommand{\lap} {$\stackrel{<}{_\sim}$}
\newcommand{\gap} {$\stackrel{>}{_\sim}$}

\newcommand{\LCDM}   {$\Lambda$CDM}
\newcommand{\etal}   {{\it et.al.}}

\newcommand{\be}{\begin{equation}}
\newcommand{\ee}{\end{equation}}
\newcommand{\mat}[1]{\ensuremath{\mathbf #1}}
\newcommand{\neff}  {$N_{\rm eff}$}
\newcommand{\mnu} {\Sigma m_\nu}

\def\taua{\mbox{Tau A}}

\newcommand{\iras}{{\sl IRAS}}
\newcommand{\firas}{{\sl FIRAS}}
\newcommand{\cobedirbe}{{\sl COBE/DIRBE}}
\newcommand{\cobedmr}{{\sl COBE/DMR}}
\newcommand{\cobefiras}{{\sl COBE/FIRAS}}
\newcommand{\so}{SO}
\newcommand{\sa}{Simons Array}
\newcommand{\pbear}{\textsc{Polarbear}}
\newcommand{\pb}{\textsc{Polarbear}}
\newcommand{\Pb}{\textsc{Polarbear}}
\newcommand{\polarbear}{\textsc{Polarbear}}
\newcommand{\act}{ACT}
\newcommand{\mrvol}[1]{\textcolor{blue}{In charge: #1}}
\newcommand{\todo}[1]{{\bf TODO: #1}}


\begin{figure}[t!]
\vspace{-.1in}
\includegraphics[width=6in]{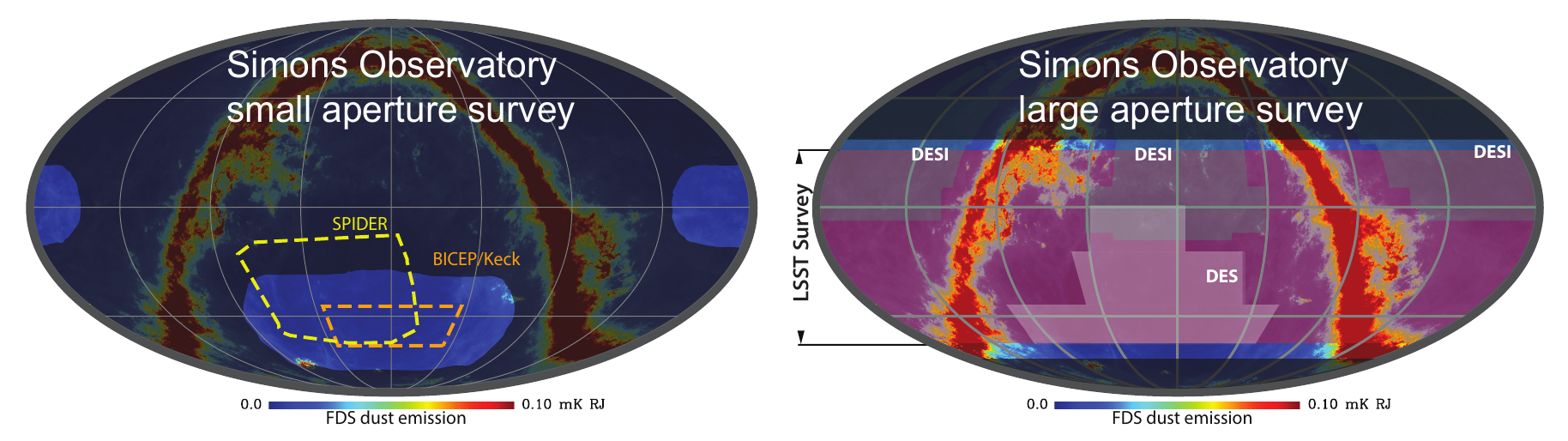}
\centering
\includegraphics[width=4.3in]{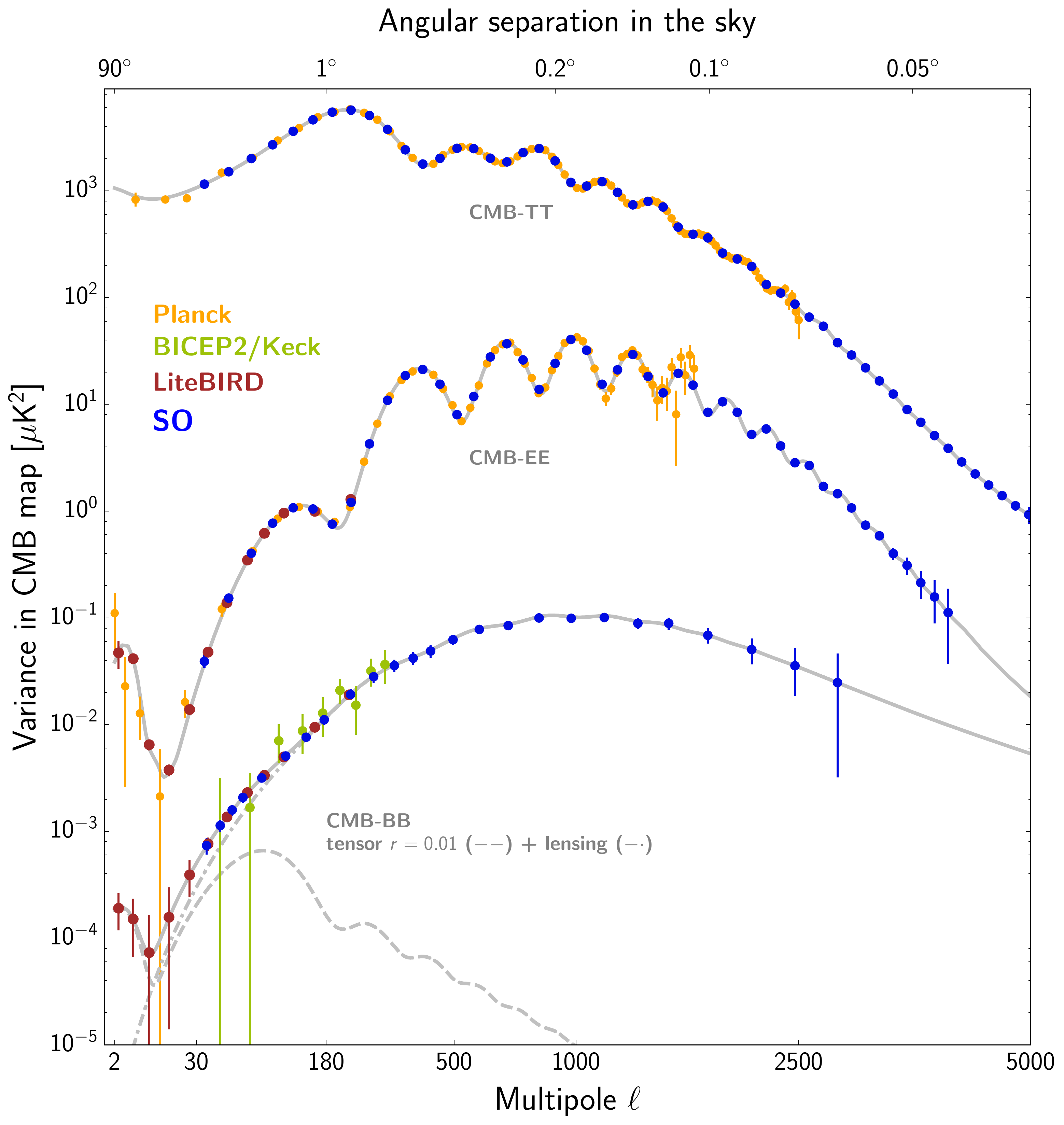}
\caption{\small (Top) Planned sky coverage of the Small Aperture Telescopes (SATs, left) and Large Aperture Telescope (LAT, right, targeting maximal overlap with LSST and DESI), in Equatorial coordinates. 
(Bottom) CMB temperature and polarization angular power spectra, showing projected SO-Nominal errors compared to current data from {\it Planck} \citep{planck_overview:2018} and the BICEP/Keck array \citep{bk:2018}, and projected errors for the LiteBIRD 0.4~m satellite. Other current ground-based data are in Fig.~18 of \cite{planck_overview:2018}. SO will increase angular resolution compared to {\it Planck}, and will improve the sensitivity of the divergence-like $E$-mode  and curl-like $B$-mode polarization signals. Other key SO statistics include the $TE$ primary spectrum, the CMB lensing power spectrum, the bispectrum, the kinematic Sunyaev-Zel'dovich (kSZ) effect, and the number of clusters seen via the thermal Sunyaev-Zel'dovich (tSZ) effect.}
\label{fig:coverage}
\end{figure}



\vskip -0.1in
This section describes the key science goals for SO-Nominal (with SO-Enhanced in \S\ref{sec:enhance}). Throughout, we indicate the relevant Astro2020 Science White Papers (SWP) with hyperlinks. Table~\ref{tab:goals} shows our key science targets for both `baseline' and  `goal' noise levels for the SATs and LAT\footnote{Parameter errors for baseline noise are increased by an additional factor of 25\%, to account for additional unforeseen noise contributions.}. To reach these targets, the LAT and the suite of SATs each have six frequency bands from 27 to 280~GHz. The SATs are expected to reach a noise level of 2$\mu$K-arcmin (baseline) and 1.4$\mu$K-arcmin (goal) coadded over the central 90 and 150~GHz channels, over 10\% of the sky where Galactic foregrounds are minimal, with half-degree resolution in those channels. The LAT should reach 6.5$\mu$K-arcmin (baseline) and 4$\mu$K-arcmin (goal) coadded over the 90 and 150~GHz channels, over 40\% of the sky where overlap with LSST and DESI is optimized, with arcminute resolution in those channels. These measurement requirements are described in \cite{so_forecast:2019}. The anticipated sky coverage and CMB power spectra uncertainties are shown in Fig.~\ref{fig:coverage}.
In the following we quote projections for baseline noise levels, with goal noise in braces \{\}.



\begin{table*}[ht!]
\centering
\caption[Simons Observatory Surveys]{Summary of SO-Nominal key science goals\textsuperscript{a}} \small
\begin{tabular}{ l |c| c  c|l|l}
\hline
\hline
  & Current\textsuperscript{b} & \multicolumn{2}{|c|}{ SO-Nominal (2022-27)} & Method\textsuperscript{d} & SWP \\
    &  & Baseline & Goal & & \\
\hline
\hline
{\bf Primordial}  &&&&&\\
{\bf \quad perturbations} (\S\ref{sec:primordial})   &&&&&\\
\hline
$r$ ($A_L=0.5$) & $0.03$ & $0.003$ & $0.002$\textsuperscript{e} & BB + external delensing & \cite{shandera/etal:2019} \\
$n_s$ & 0.004 & 0.002 & 0.002 &  TT/TE/EE &\cite{shandera/etal:2019} \\
 $e^{-2\tau}\mathcal{P}(k=0.2 \rm{/Mpc})$ & 3\% & 0.5\% & 0.4\% & TT/TE/EE & \cite{slosar/etal:2019b}\\
$f^{\rm local}_{\rm{NL}}$ &  5&  3 & 1 & $\kappa \times$ LSST-LSS & \cite{meerburg/etal:2019}\\
&& 2 & 1 &kSZ + LSST-LSS  &\\
\hline
{\bf Relativistic species} (\S\ref{sec:Neff}) &&&&&\\
\hline
\neff\ &  0.2 & $0.07$ & 0.05  & TT/TE/EE + $\kappa \kappa$ & \cite{green/etal:2019}\\
\hline
{\bf Neutrino mass} (\S\ref{sec:mnu})&&&&&\\
\hline
$\mnu$ (eV, $\sigma(\tau)=0.01$) & 0.1 &  0.04 & 0.03 & $\kappa \kappa$ + DESI-BAO & \cite{dvorkin/etal:2019} \\
&& 0.04 & 0.03 &  tSZ-N $\times$ LSST-WL &\\
&&&&&\\
$\mnu$ (eV, $\sigma(\tau)=0.002$) & &   0.03\textsuperscript{f}  & 0.02  & $\kappa \kappa$ + DESI-BAO + LB &\\
 & &   0.03  & $0.02$  &  tSZ-N $\times$ LSST-WL + LB & \\
\hline
{\bf Beyond standard } &&&&&\\
{\bf \quad model} (\S\ref{sec:de}) &&&&&\\
\hline
$\sigma_8(z=1-2)$& 7\%& 2\% & 1\% &$\kappa \kappa$ + LSST-LSS & \cite{slosar/etal:2019}\\
&& 2\% & 1\% & tSZ-N $\times$ LSST-WL & \\
$H_0$ (km/s/Mpc, $\Lambda$CDM) & 0.5 & 0.4 & 0.3 & TT/TE/EE + $\kappa \kappa$ & \cite{beaton/etal:2019}\\
\hline
{\bf Galaxy evolution} (\S\ref{sec:gf})&&&&&\\
\hline
$\eta_{\rm feedback}$ & 50-100\% & 3\% & 2\% & kSZ + tSZ + DESI &\cite{battaglia/etal:2019}\\
$p_{\rm nt}$  & 50-100\% & 8\% & 5\% & kSZ + tSZ + DESI & \cite{battaglia/etal:2019}\\
\hline
{\bf Reionization } (\S\ref{sec:reion})&&&&&\\
\hline
$\Delta z$ & 1.4 & 0.4 & 0.3 & TT (kSZ) & \cite{alvarez/etal:2019}\\
\hline
\hline
\end{tabular}
\begin{tablenotes}
\item \textsuperscript{a} Projected 1$\sigma$ errors as in \cite{so_forecast:2019}, with the addition of neutrino mass limits for an optical depth measurement of $\sigma(\tau)=0.002$, achievable with LiteBIRD soon after SO-Nominal is concluded. 
 Sec. 2 of \cite{so_forecast:2019} describes our methods to account for noise properties and foreground uncertainties. 
 A 20\% end-to-end observation efficiency is used, matching what has been typically achieved in Chile.  We assume SO is combined with {\it Planck} data.
\item \textsuperscript{b} 
Primarily from \cite{bk:2018} and \cite{planck_cosmology:2018}. We anticipate data from existing ground-based data
to improve on the `current' limits by 2022. Constraints 
are expected to lie between the `current' and SO-Nominal levels.
\item \textsuperscript{d} 
The SO observables and external datasets used, as summarized in \cite{so_forecast:2019} except for `LB' added here.
\item \textsuperscript{e} The SO-Goal projected uncertainty with zero foregrounds is 
$\sigma(r)=0.0007$.
\item \textsuperscript{f} 
Neglecting foregrounds and possible systematic errors, $\sigma(\mnu)$ would be $0.016$~eV for this case. 
\end{tablenotes}
\label{tab:goals}
\end{table*}

%


\vspace{-.2in}
\subsection{Primordial perturbations \label{sec:primordial}}
SO targets a refined theory for the early universe by searching for as-yet undetected primordial tensor perturbations, and more precisely characterizing the primordial scalar perturbations.

 \begin{figure}[t!]
\centering
\includegraphics[width=0.48\columnwidth]{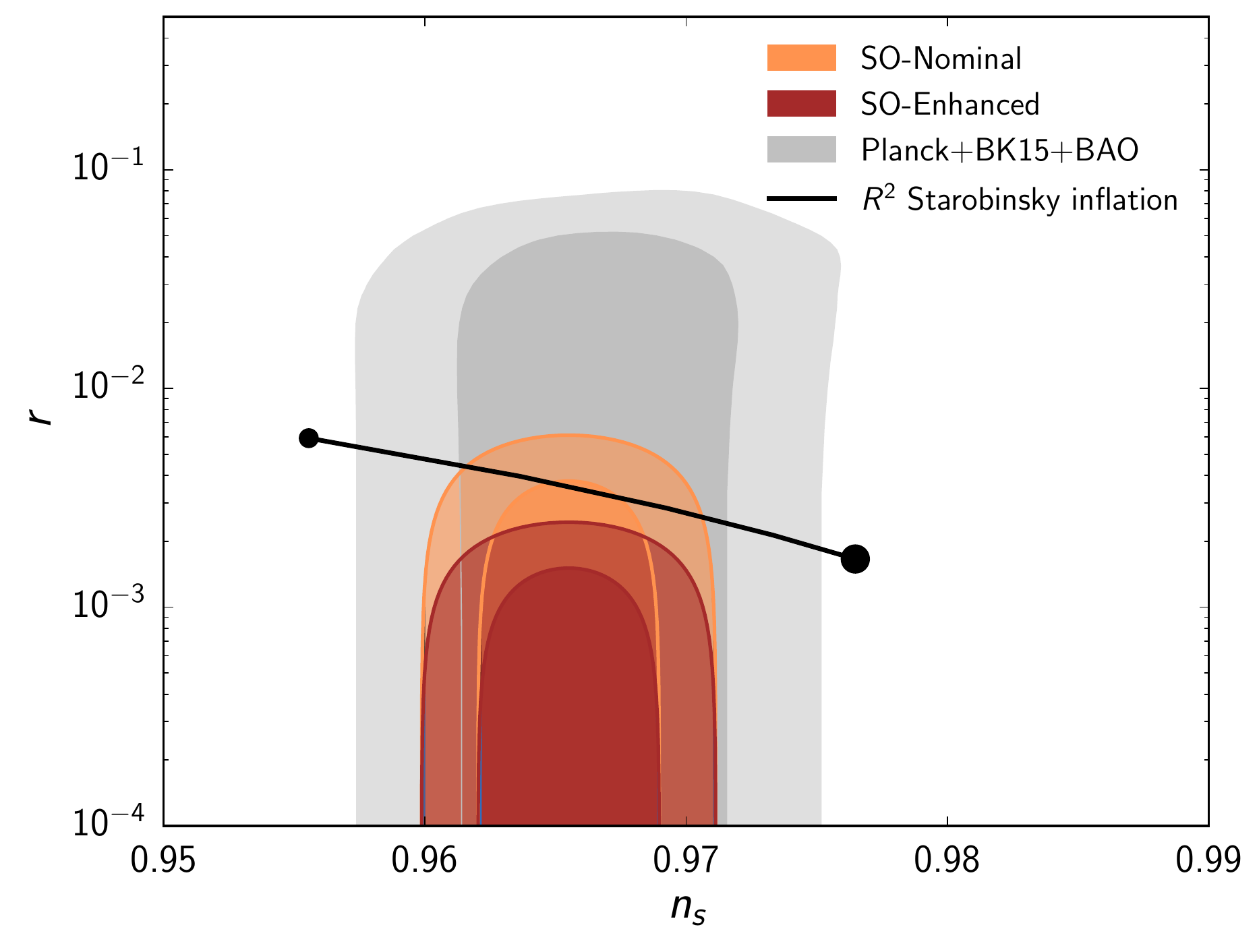}
\includegraphics[width=0.48\columnwidth]{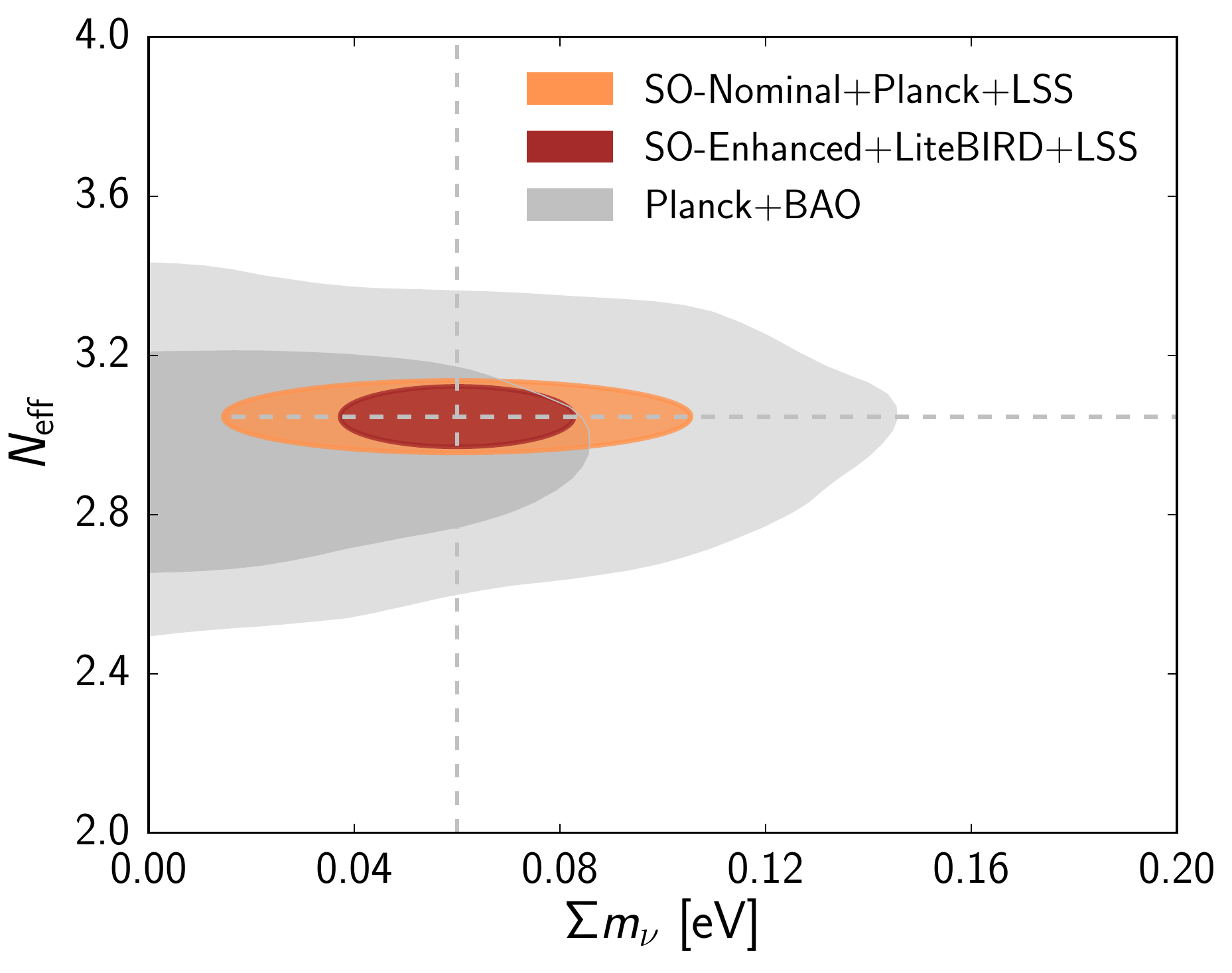}
\caption{\small {\bf Left, SO projections (68\% and 95\% limits) for primordial power spectrum parameters:} $n_s$ (the scalar spectral index) and $r$ (the tensor-to-scalar ratio) for vanishingly small tensor modes, compared to current constraints. SO-Nominal is described in  Sec. \ref{sec:science}; SO-Enhanced in Sec \ref{sec:enhance-subsec}. We expect a 3-5$\sigma$ measurement of primordial gravitational waves if $r \ge 0.01$, improving to up to 10$\sigma$ with the enhanced SO configuration. {\bf Right, SO projections for neutrino physics (68\% limits):} the sum of neutrino masses ($\mnu$) and the effective number of relativistic species (\neff). SO should give a clear indication of a non-zero mass sum in the inverted hierarchy (where $\mnu\ge0.1$~eV), and should measure the minimal normal hierarchy at $3-4\sigma$ when combined with LiteBIRD ($\mnu\ge0.06$~eV, indicated).}
\label{fig:pert}
\end{figure}

\noindent
    {\bf (a) Tensor-to-scalar ratio, ${\bf r}$} \cite[SWP][]{shandera/etal:2019}: With its SATs, SO aims to measure $r$ with $\sigma(r)=0.003~\{0.002\}$ for an $r=0$ model (Fig.~\ref{fig:pert}). 
 This will enable at least a $3\sigma$ measurement of primordial gravitational waves if $r \ge 0.01$, providing an important test of inflationary models. Otherwise, SO will lower the current limit 
by an order of magnitude. 
If no signal is seen at this level, a substantial set of inflationary models will have been ruled out.\\
 \noindent 
{\bf (b) Scalar perturbations} \cite[SWP][]{slosar/etal:2019b}: Using the LAT CMB power spectra, SO will halve the current uncertainty on the spectral index of primordial perturbations (to $\sigma(n_s)=0.002$), and will estimate the scalar amplitude at the half-percent level at $k=0.2$/Mpc scales. This will test the almost-scale-invariant prediction of inflation over a wider range of scales than accessible to {\it Planck}.
\\
 {\bf (c) Non-Gaussianity of perturbations} \cite[SWP][]{meerburg/etal:2019}: SO will constrain $\sigma(f^{\rm local}_{\rm NL})= 2~\{1\}$, more than halving current constraints.  This will be derived via either the kinematic Sunyaev--Zel'dovich effect, or the CMB lensing field, each correlated with the LSST galaxy distribution, and from the bispectrum, which will also improve constraints on other  non-Gaussianity parameters.\\


\vspace{-.4in}
\subsection{Effective number of relativistic species\label{sec:Neff} \cite[SWP][]{green/etal:2019}}
A universe with three neutrino species -- and no additional light species -- provides $3.046$ effective relativistic species. SO aims to measure $\sigma(N_{\rm eff})=0.07~\{0.05\}$, more than halving the current limit (Fig.~\ref{fig:pert}).
New light species appear in many well-motivated extensions of the Standard Model. Their contribution to $N_\mathrm{eff}$ is determined by the temperature at which they fell out of equilibrium. For example, $\Delta N_{\rm eff} \ge 0.047$ is predicted for models with additional non-scalar particles that were in thermal equilibrium with Standard Model particles at any point back to the time of reheating. 


\vspace{-.2in}
\noindent
\subsection{Neutrino Mass \label{sec:mnu}  \cite[SWP][]{dvorkin/etal:2019}}
Within the normal neutrino hierarchy the total mass in the three neutrino species has $\mnu \ge 0.06$~eV, or $\ge 0.1$~eV in the inverted hierarchy. SO aims to measure this total mass with  $\sigma(\mnu)=0.04~\{0.03\}$~eV (Fig. \ref{fig:pert}), which could give a clear indication of a non-zero mass sum in the inverted hierarchy. SO will use three complementary methods: (i) CMB lensing combined with BAO measurements from DESI; (ii) Sunyaev-Zel'dovich (SZ) cluster counts calibrated with weak lensing measurements from LSST; and (iii) thermal SZ distortion maps combined with BAO measurements from DESI. The legacy SO dataset can be combined with LiteBIRD's future cosmic variance-limited measurement of the optical depth to reionization to reach $\sigma(\mnu)$=$0.02$~eV, which would enable at least a 3$\sigma$ (5$\sigma$) measurement of the mass within the normal (inverted) hierarchy.\\

 
 \vspace{-.2in}
\subsection{Physics beyond the Standard Cosmological Model\label{sec:de}}
Upcoming optical data promise to constrain deviations from a cosmological constant at redshifts $z<1$. SO will be complementary by measuring the amplitude of matter perturbations, $\sigma_8$, out to $z=4$ with a 2\% constraint between $z=\textrm{1--2}$, and will use the kinematic SZ (kSZ) effect to enable novel tests of modified gravity theories \cite[SWP][]{slosar/etal:2019}. The matter perturbation amplitude will be obtained using SZ galaxy clusters calibrated with LSST weak lensing measurements or SO's own lensing measurements, or using CMB lensing maps cross-correlated with the LSST galaxy number density. SO's lensing measurements will also independently calibrate LSST's multiplicative bias. 

SO's precision measurement of the 
primary and lensing power spectra will enable searches for dark matter interactions \cite[SWP][]{gluscevic/etal:2019}, constraints on ultra-light axions \cite[SWP][]{grin/etal:2019},  isocurvature perturbations, cosmic strings, and precision tests of Big Bang Nucleosynthesis \cite[SWP][]{grohs/etal:2019}. SO will make the most precise measurement to date of cosmic birefringence and primordial magnetic fields. 

In looking beyond the known extensions to $\Lambda$CDM, SO will reduce the current uncertainty on the Hubble constant derived from the CMB within the \LCDM\ model, reaching a third of a percent precision measurement on $H_0$. In doing so, SO will make independent determinations of cosmological parameters using the polarization spectra (EE and TE) instead of intensity (TT), with different foregrounds and systematic effects.  A persistent discrepancy in $H_0$ values inferred from the CMB versus Hubble diagram measurements at low redshift \cite[SWP][]{beaton/etal:2019} might indicate new physics. Models that change the sound horizon in an attempt to resolve this discrepancy should be readily distinguished with the SO E-mode measurements.\\

\vspace{-.2in}
\subsection{ Galaxy evolution: feedback efficiency and non-thermal pressure in massive halos  \cite[SWP][]{battaglia/etal:2019,mantz/etal:2019}
\label{sec:gf}}
By measuring the thermal and kinematic Sunyaev--Zel'dovich effects in massive halos, using galaxy positions measured by the DESI spectroscopic survey, SO will inform and refine models of galaxy evolution. 
It will constrain the feedback efficiency in massive galaxies, groups, and clusters, $\eta_{\rm feedback}$, to 3\% \{2\%\} uncertainty, and the degree of non-thermal pressure support, $p_{\rm nt}$, to 8\% \{5\%\} uncertainty. No direct constraints on these important galaxy formation parameters have yet been derived from statistically meaningful galaxy samples. 
\\

\vspace{-.3in}
\subsection{ Reionization: measurement of duration  \cite[SWP][]{alvarez/etal:2019}\label{sec:reion}}
The reionization process is still poorly characterized. If the duration of reionization $\Delta z > 1$, SO will measure the average duration of reionization with a significance between $2\sigma$ and $3\sigma$, and probe the anisotropy of the process, thus constraining models for ionization. Such a measurement would be among the first to probe the properties of the first galaxies, quasars, and the intergalactic medium in the reionization epoch. This measurement is derived from the power spectra and four-point functions of the 
LAT maps,
since patchy reionization adds excess variance and higher order statistics to the temperature anisotropies through the kSZ effect \cite{ferraro/smith:2018}.\\

\vspace{-.3in}
\subsection{Legacy data products for science from Solar System to Cosmology}
\label{sec:legacy}


SO will produce its multifrequency maps of the microwave sky for use by the general astronomical community.  Its high level products will include  component-separated maps of the blackbody CMB temperature and polarization, the Cosmic Infrared Background \cite[SWP][]{kashlinsky/etal:2019}, and the integrated thermal pressure through the thermal SZ effect (measured to $300 \sigma$).  SO will trace the projected mass distribution with its  160$\sigma$ measurement of
gravitational lensing. Its Galactic synchrotron and dust maps will reveal insights into  
 the Galactic magnetic field \cite[SWP][]{clark/etal:2019}.


SO will produce  a legacy galaxy cluster catalog of order 20,000 clusters detected via the SZ effect, and a point source catalog of order 10,000 AGN and 10,000 dusty star-forming galaxies \cite[SWP][]{dezotti/etal:2019,dezotti/etal:2019b}.
The high resolution maps have sufficiently high mapping speed and sky coverage that they can be used to search for transient sources and/or spatially-varying sources such as Planet X and asteroids \cite[SWP][]{holder/etal:2019}.  If warranted, the scan pattern can be optimized for such searches.

When combined with complementary data from LSST, DESI, Euclid,
SPHEREx, eROSITA and WFIRST, SO will trace the evolution of the
large-scale distribution of mass, electron pressure and electron
density. 
In combination with observations from WISE, {\it Planck} CIB, and other surveys of the $z=1-2$ universe, LSST's measurements of the large-scale distribution of galaxies will provide a template for the large-scale distribution of matter.  We will use this template to undo the effects of gravitational lensing and increase SO's sensitivity to primordial gravitational waves (see $A_L = 0.3$ forecasts in Table 2).
In cross-correlations with DESI, its kinematic SZ measurements will trace the integrated electron momentum to $190 \sigma$. 

%% file: experiment.tex
\vskip -0.2in
\section{Technical Overview}
\label{sec:experiment}
The SO is designed to achieve the measurements described in Sec \ref{sec:science}, using the SO-Nominal configuration. It builds on technology and analysis techniques developed and proven by the CMB community over the last several decades.  As with previous generations, the SO has made incremental improvements based on lessons learned.
The massive scaling of the project, with five times more detectors than have ever been deployed for CMB, has required targeted significant development in 
detector multiplexing (Technology Driver, \S\ref{sec:detectors}).
Now in its implementation phase, the SO has retired most of the technical risks.  We are moving forward with an analysis pipeline design which addresses the scaling challenges associated with the large number of detectors.

The SO site at Cerro Toco is well established: CMB experiments have been operating there since 1998.  The site is near one of the largest mining centers in the world which offers a wide range of logistical advantages that reduce cost and risk.
These include year-round access via commercial airline in 1-2 days from North America, readily available commercial housing, a local skilled workforce, major construction equipment, and high-bandwidth internet allowing full data transfer and remote operation.


\vspace{-.1in}
\subsection{Small Aperture Telescopes\label{sec:SAT}}

\begin{figure}
    \centering
    \includegraphics[height=3.4 in]{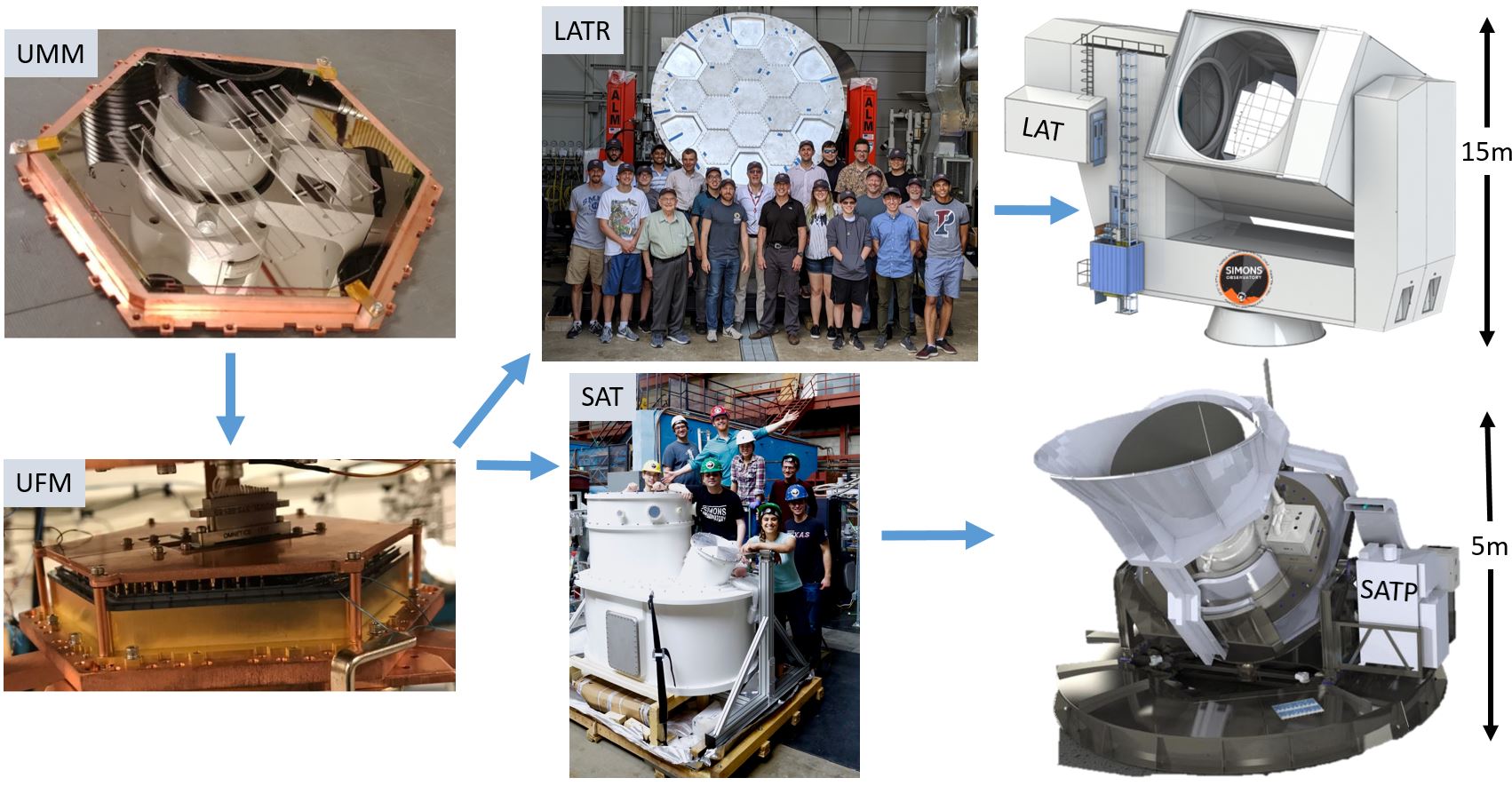}
    \caption{\small The universal microwave-multiplexing module (UMM) is combined with the detector wafer to form the universal focal plane module (UFM, shown here with dummy silicon parts).  The modules are designed such that they can be incorporated into either the Large Aperture Telescope Receiver (LATR), or the Small Aperture Telescopes (SATs).  The LATR is then installed in the Large Aperture Telescope (LAT) and the SAT is installed on the SAT Platform (SATP).}
    \label{fig:Instrument}
\end{figure}

The three SATs of SO-Nominal, each with a single optics tube, will together contain 30,000 detectors operating at 100~mK. They adopt a design highly optimized for measuring the polarization of the CMB at degree scales (Fig. \ref{fig:Instrument}). Each SAT houses seven detector arrays, and will have a continuously rotating half-wave plate to modulate the polarization signal. 
Two SATs will observe at 93 and 145 GHz (Mid-Frequency, MF), one at 225 and 280 GHz (Ultra-High Frequency, UHF); an additional low-frequency optics tube at 27 and 39 GHz (Low Frequency, LF) will be deployed for a single year of observations. 

The optics system has a 42-cm stop aperture and a 35-degree diameter field of view.  This is enabled by three 46-cm diameter silicon lenses with anti-reflecting metamaterial surfaces.  The entire optical system is cooled to 1~K. 
An intensive ring-baffle scheme with engineered black body absorbers controls stray light.
A 50~cm diameter rotating 40~K sapphire half-wave plate is located near the sky side of the aperture stop.
This significantly mitigates instrumental systematic effects and rejects atmospheric fluctuations, crucial for sensitivity to degree-scale polarization.

Warm optical baffling comprises three stages: a forebaffle attached to the cryostat, co-moving shield attached to the elevation stage of the platform, and a fixed ground screen.
Electromagnetic simulations verify the effectiveness of this multilevel approach for the control of far sidelobe and ground-synchronous systematic effects.
The entire SAT cryostat, forebaffle and co-moving shield, and the supporting electronics are mounted on a three-axis platform.  The platform allows fast scanning in azimuth, elevation, and rotation around the line-of-sight for systematic control.


\vspace{-.2in}
\subsection{Large Aperture Telescope \label{sec:LAT}}

The LAT is a 6~m diameter crossed-Dragone telescope designed with an 8$^\circ$ field of view providing roughly seven times the throughput of current Stage 3 telescopes (Fig. \ref{fig:Instrument}). The LAT is being built by Vertex Antennentechnik 
in parallel with the CCAT-prime sub-millimeter telescope, and both are planned for installation in Chile in 2021 \citep{parshley/etal:2018}.

Designed to take advantage of the capabilities of the LAT, the LAT receiver (LATR) will be the largest cryogenic camera for CMB studies ever built.  
It is 2.5~m in diameter and over 2.6~m long accommodating up to 13 optics tubes.  It must cool a metric ton of material to 4 degrees above absolute zero and over 100 kg to 0.1 degrees above absolute zero.  

The LATR is designed to maximize sensitivity while implementing new designs to increase robustness against systematic effects.  
Each of the 13 optics tubes consists of three state-of-the-art anti-reflection coated silicon lenses and an array of filters.  The optical design enhances control of stray light which could 
limit the sensitivity of the instrument.  Extensive modeling based on past experience has driven novel design changes to meet the new specifications.  The LATR will be deployed with 7 of the 13 optics tubes, comprising SO-Nominal (with 1 LF, 4 MFs and 2 UHFs). The additional 6 tubes can be added in the field, and are part of the SO-Enhanced program (\S\ref{sec:enhance}).


\vspace{-.2in}
\subsection{Detectors and Readout\label{sec:detectors}}




The focal planes of each of the SO telescopes will be tiled with hexagonal universal focal-plane modules (UFMs, Fig.~\ref{fig:Instrument}) based on dichroic transition edge sensor (TES) bolometric polarimeters. Each pixel contains four bolometers:  two frequency bands in two orthogonal polarizations.  The detector arrays for SO come in three frequency configurations.  The MF (90/150 GHz) and UHF (220/270 GHz) UFMs host $\sim 2000$ bolometers each, while the LF (27/40 GHz) UFMs have $\sim 200$ each. SO will employ bolometer arrays coupled through orthomode transducers to aluminum feedhorns, and arrays coupled through dual-polarization sinuous antennas and lenslets. These two flavors of arrays are made at NIST and UC Berkeley respectively, giving the project built-in risk mitigation against schedule delays due to production hold-ups at one fabrication house.   

{\bf Technology driver:} 
The bolometers are read out with a microwave multiplexing scheme \citep{mates/etal:2017,dober/etal:2017} that transduces each detector's CMB signals into frequency shifts of a cryogenic resonator tuned between 4 and 8 GHz. Here 1800 resonators are coupled to a single transmission line so that one UFM (or up to seven LF UFMs) is read out with a single pair of coaxial cables and a dozen pairs of lines carrying biases for the detectors and readout. For efficient filling of the focal planes, each UFM packs the cryogenic readout components directly behind a detector array fabricated on a 150 mm wafer. The cryogenic readout components for a UFM are collectively called the universal microwave-multiplexing module (UMM), which includes a few dozen SQUID-based multiplexing chips and silicon boards providing shunt resistors and series inductors for each TES, mounting, and routing of lines. Each UMM is validated as a unit, and then coupled to a detector stack comprising a detector wafer with associated optical coupling components on its sky side and back-termination components on its non-sky side. The resulting UFMs are validated three at a time in one of several 100-mK test systems using a combination of dark tests and cryogenic load tests. The multiplexing method has been demonstrated recently with 528 resonators \cite{henderson/etal:2018} but requires maturation to scale it to 1800 and adapt it to the necessary layout for SO.  We have fabricated and assembled prototype modules at full scale in the appropriate layout and are currently testing them.